**Bridging microscopy with molecular dynamics and quantum simulations: An AtomAI based pipeline**

Ayana Ghosh,[1,2,a] Maxim Ziatdinov,[1,2] Ondrej Dyck[1], Bobby Sumpter[1], and Sergei V. Kalinin[1]

[1] Center for Nanophase Materials Sciences, Oak Ridge National Laboratory, Oak Ridge, TN 37831

[2] Computational Sciences and Engineering Division, Oak Ridge National Laboratory, Oak Ridge, Tennessee 37831, USA

Recent advances in (scanning) transmission electron microscopy have enabled routine generation of large volumes of high-veracity structural data on 2D and 3D materials, naturally offering the challenge of using these as starting inputs for atomistic simulations. In this fashion, theory will address experimentally emerging structures, as opposed to the full range of theoretically possible atomic configurations. However, this challenge is highly non-trivial due to the extreme disparity between intrinsic time scales accessible to modern simulations and microscopy, as well as latencies of microscopy and simulations per se. Addressing this issue requires as a first step bridging the instrumental data flow and physics-based simulation environment, to enable the selection of regions of interest and exploring them using physical simulations. Here we report the development of the machine learning workflow that directly bridges the instrument data stream into Python-based molecular dynamics and density functional theory environments using pre-trained neural networks to convert imaging data to physical descriptors. The pathways to ensure the structural stability and compensate for the observational biases universally present in the data are identified in the workflow. This approach is used for a graphene system to reconstruct optimized geometry and simulate temperature-dependent dynamics including adsorption of Cr as an ad-atom and graphene healing effects. However, it is universal and can be used for other material systems.

[a] ghosha@ornl.gov

**Introduction**

Over the last two decades, electron and scanning probe microscopies have evolved as one of the primary tools to study systems in the domain of physical and life sciences at the atomic to mesoscopic length scales.[1-5] Advances in (Scanning) Transmission Electron Microscopy, (S)TEM, and scanning tunnelling microscopy (STM) measurements produce highly reliable structural and spectral data containing a wealth of information on structures and functionalities of materials. In particular, t aberration corrected STEM inclusive of single-atom EELS imaging allows study of single impurity atoms,[6] structures with grain boundaries,[7] probe orbital,[8-9] and magnetic phenomena,[10] plasmons,[11-12] phonons,[13] and even the anti-Stokes excitations in complex materials.[14]

There also exists ample detailed fundamental studies exploring quantum corrals,[15] molecular cascades,[16] quantum dots along with investigations of surface chemistry[17-18] that have been carried out utilizing STM measurements. STEM offers the potential for much higher throughput imaging and data generation as compared to STM due to intrinsic latencies in electron beam motion and image acquisition. Both in STM and STEM, the probe can induce changes in materials structure. In STM, this is often associated with probe damage as well, and in most cases, perceived as a negative effect; controllable modifications of surface by an STM probe are actively pursued for atomic fabrication and exploration of surface chemistry.[16, 19-21] Comparatively, in STEM the changes in material structure do not affect the probe state, rendering this technique a powerful tool for exploring metastable chemical configurations and beam-, temperature-, and chemistry induced processes.[22-25] It is possible to manipulate atoms[26-27] and corresponding positions and thereby solids and molecular systems with both techniques in controlled environments.

The images generated at different stages of STEM observations contain a wealth of information of materials structures, functionality, and chemical transformation pathways encoded in observed positions. The learnings from such experiments can be both qualitative and quantitative in nature. Images generated by such observations lead to high-resolution data, creating a platform to build deep learning (DL) models for finding features,[28-29] predicting scalar functional quantities (such as values of ferroelectric polarizations) or even crafting chemical or structural space maps. Comprehensive studies[30-36] utilizing such datasets combined with modern DL techniques such as convolution networks, variational autoencoders as implemented in general machine learning (ML) frameworks[37-39] or ensemble settings, have already shown great potentials to advance physics-based understanding of materials by establishing causal relationships between structures and properties.

On the other hand, development and availability of more computational capabilities including accessible CPU/GPUs, efficient algorithms and corresponding implementations have significantly boosted the advancement of physical simulations. Physical models constructed using first-principles theory to quantum Monte Carlo (MC) and finite-element methods, spanning over quantum-mechanical to continuum scales, lead to an abundance of insights on structural, thermodynamic, and electronic properties of solids, glasses, and liquids. In general, atomistic simulations provide information on atomic length scales where continuum theory breaks down and instead complex many-body quantum-mechanical theory comes into play to model behavior of each atom and how collectively these atoms give rise to properties of a material under specific conditions. Even though such detailed studies have become quite common in the past decade, atomistic understandings do not readily translate to the macroscales, and hence there has been a consistent effort in the scientific community to merge multi-scale investigations performed across overlapping length and time scales. If we compare some of the nuances of various simulations, we can easily narrow down the primary challenges and the need for transferring the knowledge between each type of these models. For e.g., density functional theory-based (DFT) computations lead to studying behaviors of materials conventionally at 0K and molecular dynamics (MD) techniques can overcome such challenge by including temperature-dependent comportment. The accuracy of MD simulations is often limited by type of particle-particle interactions represented by force-fields as well as the dilemma to choose between computation efficiency and length of simulations. Consequently, a slow thermodynamic process like diffusion cannot always be modeled by this approach and that is where MC methods can become useful. It is possible to randomly probe molecular systems using such simulations, enabling researchers to study various mechanisms, steady state properties and even dynamics using kinetic MC. There also exists several advanced frameworks combining some of these methods such as ab-initio MD (AIMD) that integrates quantum-mechanical estimation of interatomic forces and classical Newtonian physics to move atoms from one instant to another. Overall, irrespective of the simulations utilized to study complex behavior of material systems, there is always a colossal flow of data that gets generated. This is one of the many reasons why data-driven and ML approaches have become so popular in recent years in any scientific domain.

There is a growing availability of databases[40-45] collating simulations and experimental data across disciplines, which are being employed to accelerate the discoveries of novel materials and study advanced material functionalities. A variety of successful examples[46-63] are illustrated for technological applications in the fields of energy, catalysis, and photovoltaics, for pharmaceutical applications in drug design and reaction mechanisms mapping, as well as in advancing fundamental knowledge of materials behavior, including magnetism, ferroelectricity, and superconductivity. There are also exciting studies showing the efficacy of ML and DL techniques to facilitate meaningful contributions to solve electronic structure, force-field related technical challenges. Most of these constructed frameworks either are solely built on theoretical, simulations-based, or experimental data or at times combinations of these. In addition, there are examples of direct comparisons of endpoint-like properties such as polarization, magnetic structures obtained from simulations and measurements that can be found in literature.

In addition, there are already comprehensive efforts in the materials community to bridge the gap between knowledge acquired from experiments and theory, to go from observational to synthetic learning and vice versa. A non-exhaustive list of such frameworks include Ingrained[64], EXSCLAIM[65], BEAM[66], abTEM[67] show how to utilize already existing data from the literature to create labeled datasets, use image-based data and parameters based on users' choice to find optimized fit-to-experiment structures via forward modeling, perform scalable data analyses and simulations on characterization data, compute potential via DFT, simulate standard imaging modes. In addition there is an extensive list of examples from the STEM to model electron beam effects for various materials systems including atoms assembly, atomic manipulations or insertions.[68-71] Finally, multiple reports on 2D materials are available in the literature that have explored formations, dynamics and stability of defects, edge reconstructions, bond inversions using combination of STEM observations and simulations.[72-82]

However, systematic studies utilizing an across-the-board framework mapping directly between experimental observations and computational studies using DL approaches is still in its infancy. As a first relevant aspect, the time scales of STEM observations and intrinsic molecular dynamics are strongly different, with the DFT and MD models capable of simulating system sizes of Å to nanometer scales for up to microseconds, but taking multiple CPU hours, while STEM images are typically available at the fraction of a second. The length and timescales of such simulations, analyses methods and retrieval of observational data are shown by a comparative chart in Figure 1. At the same time, there is significant disparity in the latencies of calculations, with the DFT or MD simulations often taking many hours to days and weeks of time, well above the timescale of STEM measurements.

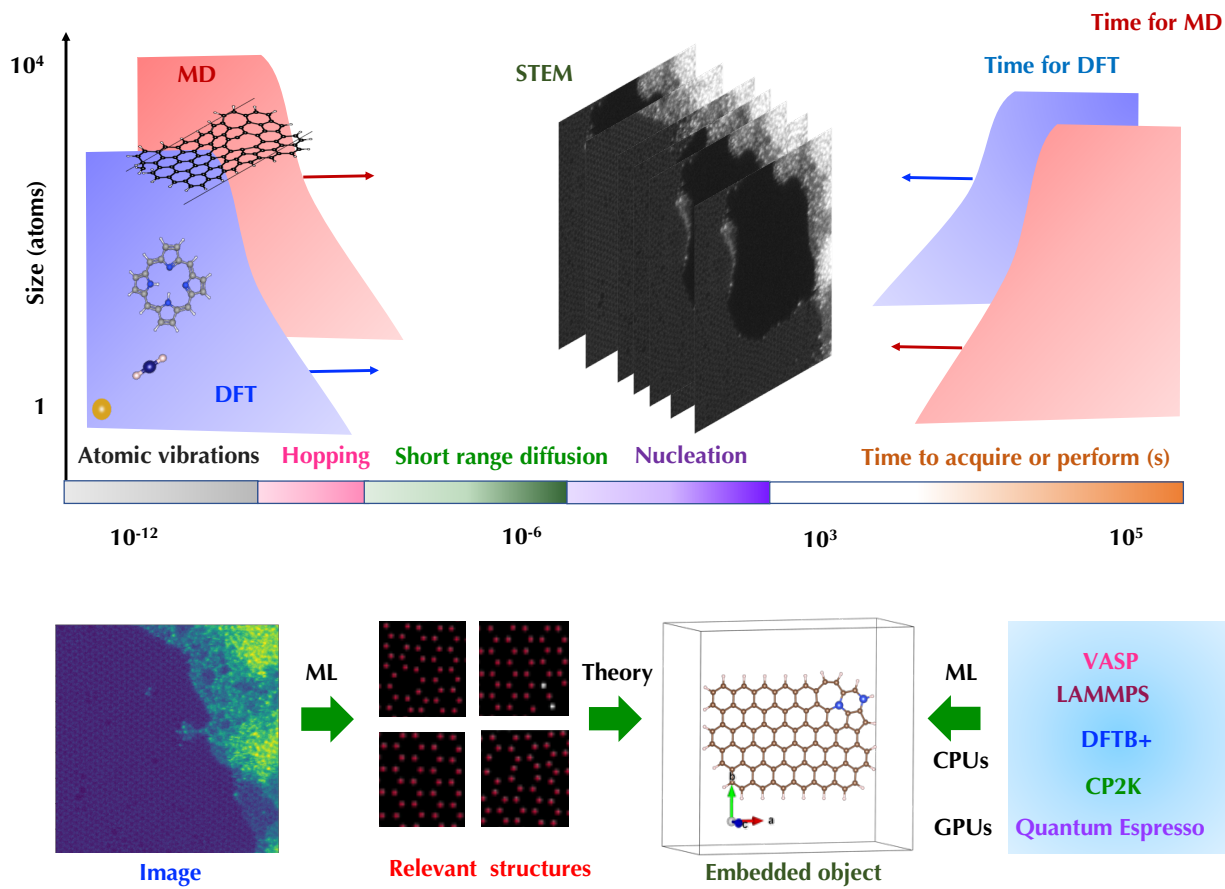

**Figure 1**: **The time and length scales for imaging and simulations**. While DFT and MD can address the system dynamics on the picoseconds to microseconds, electron microscopy generally explores dynamic changes within milliseconds – hours interval. This significant time disparity is further exacerbated by the fact that the time to complete DFT/MD simulations for a sufficiently large system is typically hours or days, way above the imaging time. Hence, dynamic integration of theory and experiment in single workflow necessitates the use of machine learning both to identify (based on observations) the regions of potential interest and to accelerate simulations via proxy models, and theory to define proper embeddings that preserve the functionality of material in small number of atom models.

Hence, while the theoretical methods directly match the experimental observables, there is drastic mismatch between both the accessible and computational timescales, making the integration between the two highly non-trivial and necessitating development of strategies to deal with this mismatch and, importantly, formulating the physics-relevant questions that can be addressed.

With these caveats, the first enabling step towards bridging these two areas is direct piping of the STEM data into the simulation environment. The crucial roadblocks to materialize this framework as listed below.

- o Finding features such as atoms, defects (nearly identical objects) from a microscopic image using DL model and extending it to recognize features from images retrieved under different experimental conditions leading to out-of-distribution effects, is itself a challenge.
- o Defining regions of interest such as parts of the image showing defects and determining the origins of such defects (could be electron-beam induced) is also not trivial. This gives rise to a choice of an intractable number of possibilities to define the initial states of the simulations.
- o Importing coordinates of atoms directly predicted by either a DL model or experiments to simulations require quantifying uncertainties at all stages of the framework. In other words, predictions with high uncertainty, if transferred to initiate simulations, it may never converge.
- o Ways to close the loop where information from the theoretical simulations can guide future experiments and on-the-fly analyses, are also yet to be investigated.

In this work, we show how deep learning can bridge together the knowledge learned from microscopic images (stage 1 in Figure 2) and first-principles simulations to develop a comprehensive understanding of the physics (stages 2 and 3 in Figure 2) of the materials of interest. The schematic of the entire workflow is shown in Figure 2. Here, we focus on how deep convolutional neural networks can be employed to identify atomic features (type and position) in graphene, use them to construct supercells, perform DFT simulations to find optimized geometry of the structures followed by studying temperature-dependent dynamics of system evolutions with ad-atoms and defects. The results along with associated uncertainties in predictions at various levels as obtained utilizing this framework may be used to evaluate and modify experimental conditions and regions of interest.

**Results and Discussion**:

*Workflow and its implementations*

The first stage of the workflow involves application of DL to experimental imaging of atoms. A standard DL workflow consists of preparing a single labeled training set, choosing suitable neural network architecture, dividing the prepared training set into training, test, and validation sets, and tuning the training parameters until the optimal performance on the test and/or 'holdout' set is achieved. Once the labelling is accomplished, the DL models are utilized to find the features, in this case, positions and types of atoms (C), defects (Si) and the uncertainties are determined in Stage 2 using AtomAI.[37]

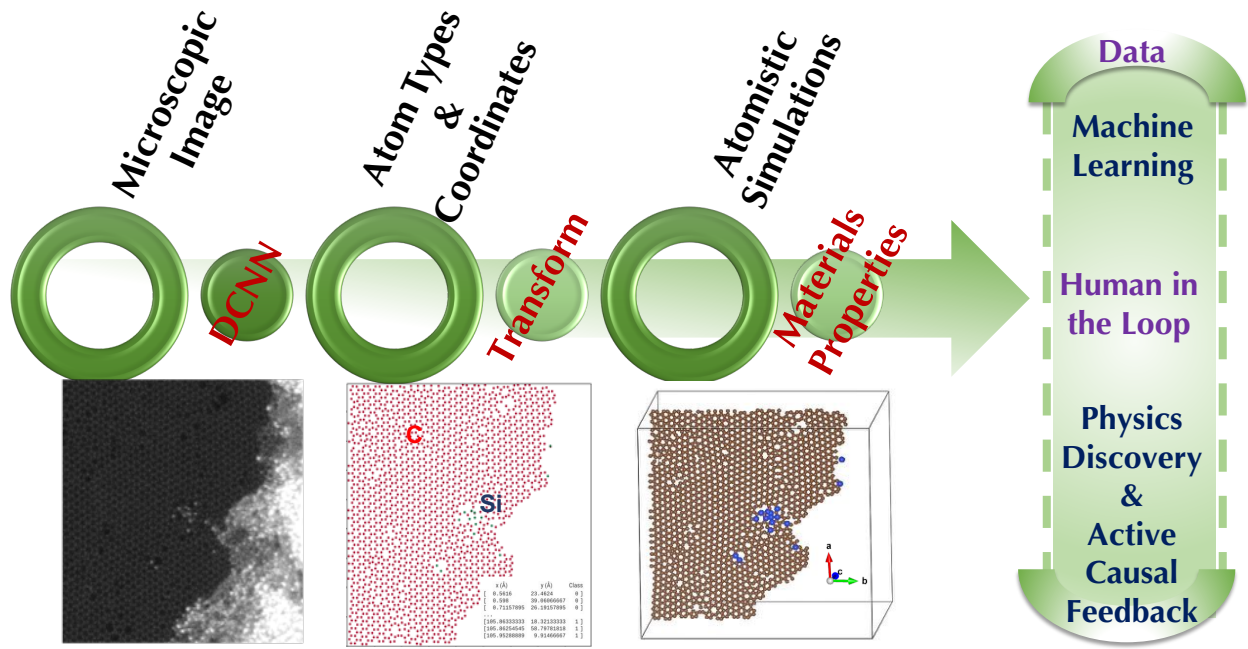

**Figure 2**: **Schematic of the workflow, from images to evaluating material properties**. Figure shows three primary stages of the framework. Stage 1 consists of how deep learning models take microscopic images (STEM image of graphene) as inputs and identifies features such as atoms, defects, and respective positions. In Stage 2, the coordinates are put together to build a simulation object followed by performing atomistic simulations in Stage 3 to study physical phenomena. The results from such simulations are later used to better build and guide physics-informed experiments.

Figure 3 shows one of the STEM graphene image frames (a) and corresponding C and Si atoms, as predicted by the DL (b) model. We note that the associated uncertainties in such predictions may vary from one image frame to another. With minimal to null human interventions, it is also possible to make sure that all atoms are identified accurately. However, the focus of this framework is to enable the transformation from microscopic image to ML predictions directly (automated) to simulation environments. As a part of Stage 3, simulation objects (could be bulk, supercell, surface) are created (c) using AtomAI utility functions. These objects are constructed such that the simulation cells (cubic or any Bravais lattice type of user's choice) can accommodate all atoms with acceptable bond lengths and imposed periodic boundary conditions, as recognized via the specific sample view or perspective.

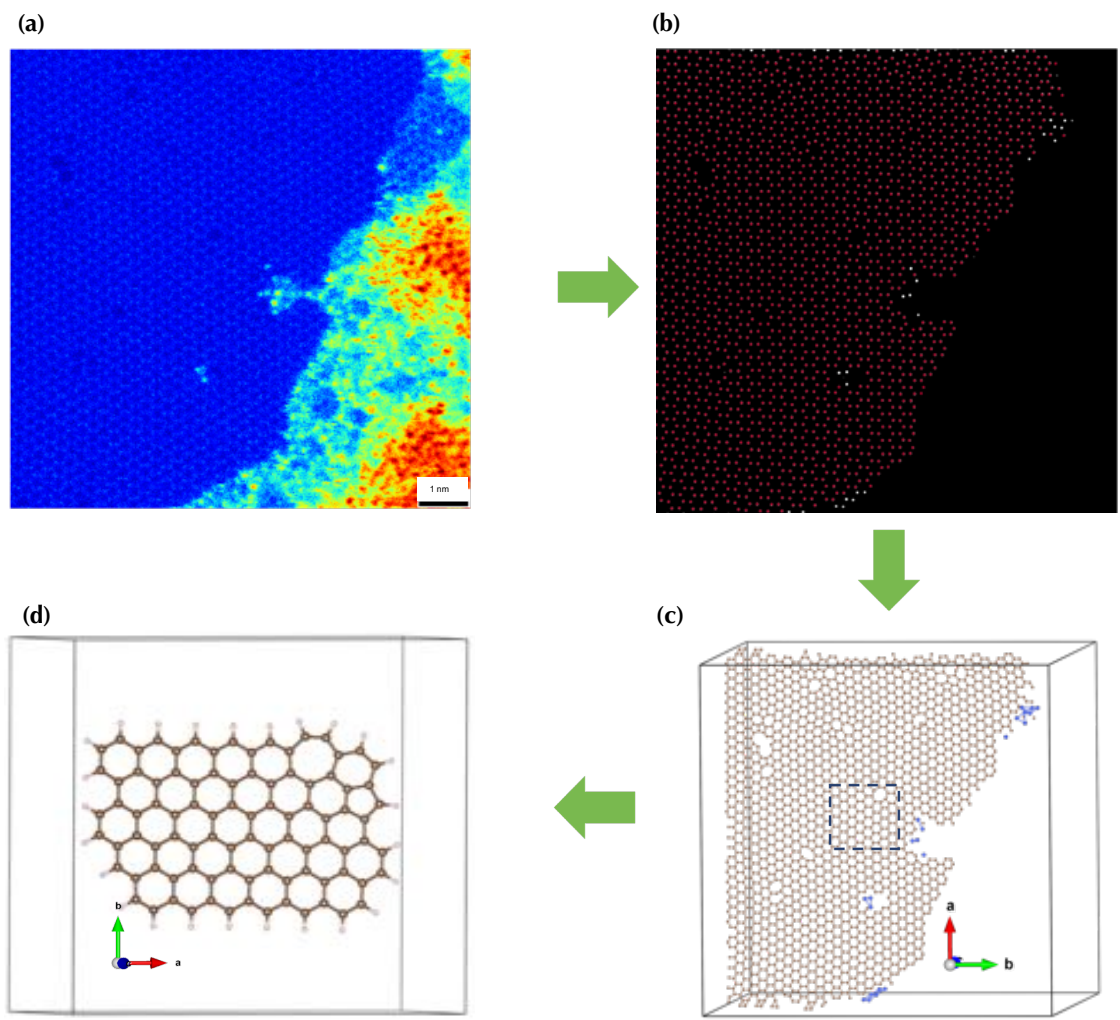

**Figure 3**: **Obtaining simulation boxes from STEM image**. Figure shows a STEM image frame (a) of graphene, all C atoms, and defects (Si) as recognized by DCNN (b), corresponding simulation object (c) and an example patch (identified region of interest) used (d) to perform DFT and AIMD simulations. The blue atoms represent Si atoms, and the brown ones are C atoms. Here, we note that some atomic species got misclassified but it did not affect the workflow since we cropped the high-certainty (and correctly classified) region. This is somewhat expected during the actual real-time deployment of the workflow.

We note that for this image frame, there are a total of 2021 C and 22 Si atoms that were detected. While performing DFT computations using these many atoms is possible with the advancement in parallelized codes and computational power, it is heavily dependent on availability of such resources. Furthermore, it may not even be interesting from the physics point of view to explore the full structure. As an alternative, we can identify several parts or patches such as those containing defects using our domain knowledge and explore these regions of interest. In addition, the system is continuous in one image frame, meaning the graphene sheet spans over the full frame, be it in lattice or amorphous phase. However, for simulations, we only assume the lattice phase and the rest to be vacuum corresponding to the amorphous region within the DFT framework. Hence, we terminate all 2-fold coordinated C atoms with H for any 'cropped' structure, irrespective of whether they are surrounded by graphene lattice or amorphous phase in the original structure to maintain charge neutrality and stoichiometry. Figure 3(d) shows one of the identified patches that is used to perform simulations to obtain optimized geometry and investigate electronic properties and temperature-dependent dynamical evolutions.

*Details on DL models*

For DL-based image analysis, individual images are labelled at the pixel level, where each pixel represents an atom, impurity, or a background. Recognizing features as represented by individual pixels is referred to as semantic segmentation. This differs from a typical image-classification task of natural images where the image gets categorized as a whole. Each graphene image frame is of dimension (height = 896 pixels * width = 896 pixels) and 100 such frames combined are used to construct the training set. Each pixel is of 0.104 Å length as considered to convert pixels into cartesian coordinates based on the STEM scan size. A U-Net [83] type neural network as used in this workflow takes the images as the input and gives output as clean images of circular-shaped "blobs" on a uniform background, from which one can identify the (x,y) centers of atoms. At the initial stages of the DL workflow, atoms are classified based on the variation in intensity using Gaussian mixture model. In the later stage of the training, such information is utilized to generate multi-label segmentation mask to perform multi-class classifications for both type of atoms, as present in the system. Here, the predictions are multi-class representing the C and Si atoms. We have utilized an ensemble setting[84] such that an artifact-free model or a subset of models can be identified that is capable of not predicting any 'unphysical features' in experimental data and can be extended to provide robustness and pixel-wise uncertainty estimates in prediction as one transitions from one experiment to another. The uncertainty is estimated as a standard deviation of ensemble predictions for each pixel. For our predictions within the DL framework, we can obtain spatial maps of uncertainty estimates. The regions characterized by high uncertainty may be due to some "unknown" defects/species (that were not a part of training data). This may lead to further investigation of those specific regions closer and explore them with DFT/MD (e.g., by adding functional groups, etc.). This procedure can be repeated multiple times to achieve a high detection rate for the entire dataset of dynamical data. In addition, it is important to note that the DL framework exploited here may also help to minimize the effects associated with the out-of-distribution effects[85] as present in the observational space due to variations in experimental parameters. The utilization of atom positions with high confidence level is crucial before transforming these predicted DL coordinates to simulation objects such that a meaningful initial state for each simulation is guaranteed.

*Geometry reconstruction using first-principles computations*

All first-principles computations are performed using DFT within the Generalized Gradient Approximation (GGA) framework and more details on the computations are given in the Methods section. To generalize this framework and make it more open-access, we first explored the possibilities utilizing Python-based codes which consist of computationally inexpensive algorithms to optimize electronic structure and consequently use the resulting system to perform MD simulations. More details on this implementation can be found in the associated Jupyter notebook given in the 'Code Availability' section. It is important to mention that appropriate reconstruction of geometries of such structures cannot be obtained using methods like quasi-Newton algorithms and pair-potentials to study dynamics, as implemented in the popular Python-based frameworks.[86-87]

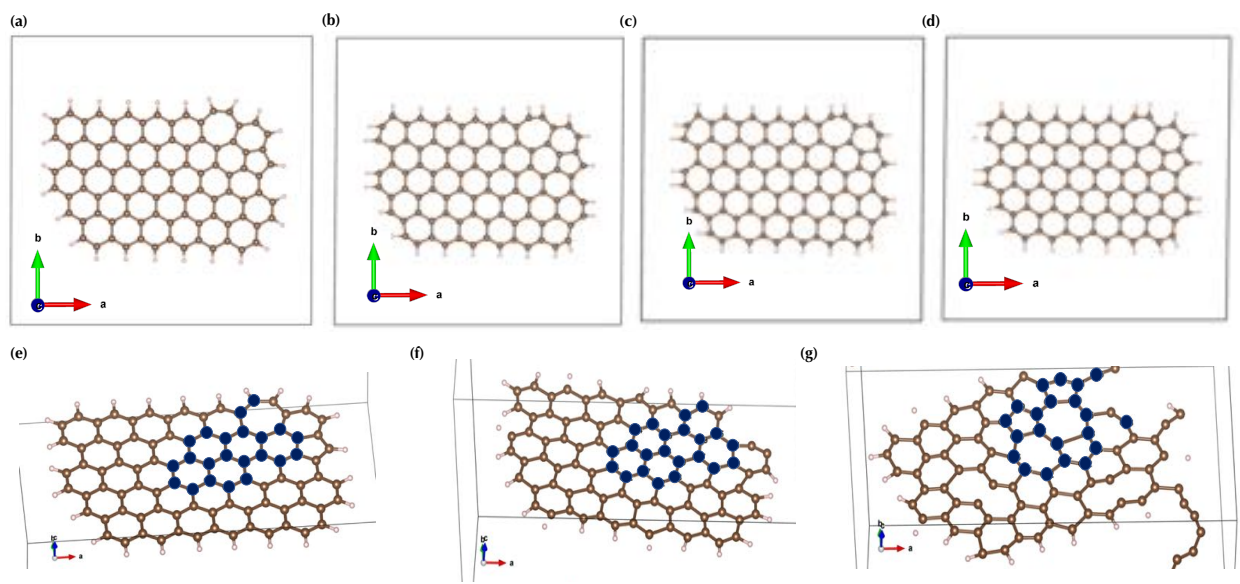

**Figure 4**: **Optimized geometries obtained using DFT within GGA computations and AIMD simulations**. Figure (a) shows the fully relaxed structure while (b-d) are the resulting optimized geometries using three different selective dynamics configurations. Figure (e-g) represent the resulting structures of graphene after 2 ps at 300 K, 2000 K and 4000 K, respectively.

Hence, we have employed DFT within GGA using VASP to first obtain the optimized structures. Once the forces and stress tensors are minimized within a given criteria by solving the Kohn-Sham (KS) equations, the optimized geometry is imported to the AIMD environment. DFT has been known for yielding reasonable geometries, meaning lattice parameters, bond angles, bond lengths along with coordinates of atoms for a bulk or surface cell can be comparable to that retrieved from experiments. We note that more detailed computations techniques such as coupled cluster, DFT with tight binding and even DFT with inclusion of van der Waals interactions, would be more appropriate to include dispersions that are important for graphene surfaces. However, these are computationally demanding and goes beyond the scope of this paper that is focused on establishing a workflow, going from a microscopic image to performing simulations with recognized features with the help of DL approaches.

Here, we started by constructing a graphene supercell (patch extracted from the full 2,000 atoms supercell) of 91 C atoms with lattice parameters a = b = c = 25.786 Å, with $\alpha = \beta = \gamma = 90°$, where all the 2-fold C atoms are terminated with H atoms. Along with aiming for the full geometry optimization where all atoms (C and H), cell volume, and shape are in the ground-state, we also designed different configurations to perform selective dynamics within DFT simulations. These include computations where (a) all the perimeter atoms positions are fixed, (b) coordinates in a, b or (x and y) directions are fixed, only coordinates along z are allowed to relax and (c) positions of all H-atoms remain fixed. The resulting structures are shown Figure 4 (a-d). The average C-C and C-H bond lengths resulting from the full geometry optimizations corresponding to Figure 4(a) are 1.42 Å and 1.09 Å, respectively. For the simulations performed with selective dynamics, these bond lengths for C-C and C-H vary between 1.43 - 1.50 Å and 0.94 - 1.04 Å, respectively. Such changes can be attributed to the change in the charge densities of the atoms, especially around the already existing defect regions. [88-89] We do observe a potential for formation of a carbon cluster defect region that was proliferated in the initial state (may have formed during growth or could be electron beam induced). The structures are all metallic in nature, as expected. Whether this region may show and evolve into one of the common defects such as Stone-Wales defect (5-7-7-5) or 5-8-5 defects at higher temperatures, that can be explored using AIMD simulations under various annealing conditions or environments. If we compare the coordinates in x, y and z directions for all C atoms as obtained after DFT simulation and those predicted by DL, the average percentage error in the coordinates along x and y directions are <5% which is reasonable. The full list of coordinates from both predictions along with point-by-point errors can be found in Table 1 of Supplementary Material. Overall, the optimized geometries of this representative patch can be reconstructed very well from starting with the ML-predicted coordinates using DFT-based computations as shown by this stage of the workflow.

# Temperature-dependent system evolutions using AIMD simulations

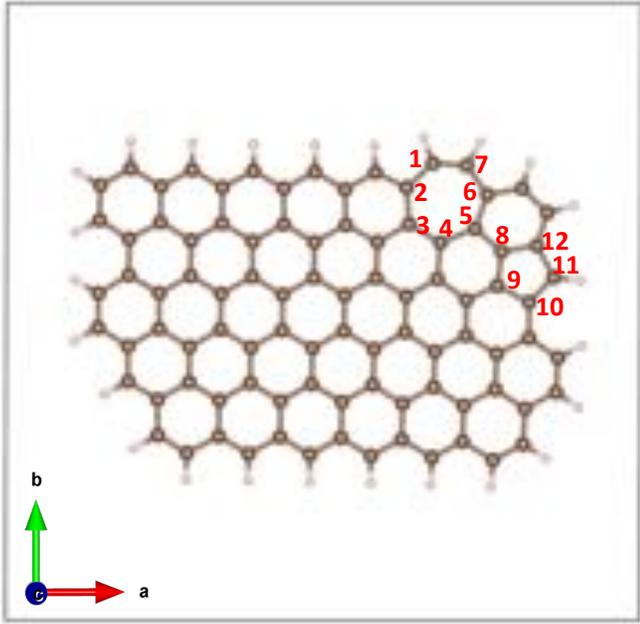

Structure with C atoms in the heptagon and pentagon rings marked

| Temperature (K) | C$_{1-2}$ (Å) | C$_{2-3}$ (Å) | C$_{3-4}$ (Å) | C$_{4-5}$ (Å) | C$_{5-6}$ (Å) | C$_{6-7}$ (Å) | C$_{7-1}$ (Å) |
|---|---|---|---|---|---|---|---|
| 300 | 1.473 | 1.478 | 1.447 | 1.509 | 1.454 | 1.427 | 1.344 |
| 1200 | 1.422 | 1.487 | 1.455 | 1.504 | 1.457 | 1.412 | 1.372 |
| Temperature (K) | C$_{8-9}$ (Å) | C$_{9-10}$ (Å) | C$_{10-11}$ (Å) | C$_{11-12}$ (Å) | C$_{12-8}$ (Å) | | |
| 300 | 1.400 | 1.444 | 1.396 | 1.451 | 1.438 | | |
| 1200 | 1.409 | 1.420 | 1.434 | 1.424 | 1.458 | | |

**Table 1**: Distances between C atoms in heptagonal and pentagonal rings at 300 K and 1200 K temperatures after 2 picoseconds.

To study the temperature dependent behavior of the fully relaxed system obtained from DFT at 0K, we move on to applying AIMD within VASP. The temperatures considered using a Nose-Hoover thermostat to control the heat bath are 300K, 500K, 700K, 900K, 1200K, 2000 K, 3000 K, 4000 K and 5000 K. Here, the simulations are performed for 2 picoseconds and rest of the computational details can be found in the Methods section.

| Structure with C atoms in the heptagon and pentagon rings marked ||||||||
|---|---|---|---|---|---|---|---|
| 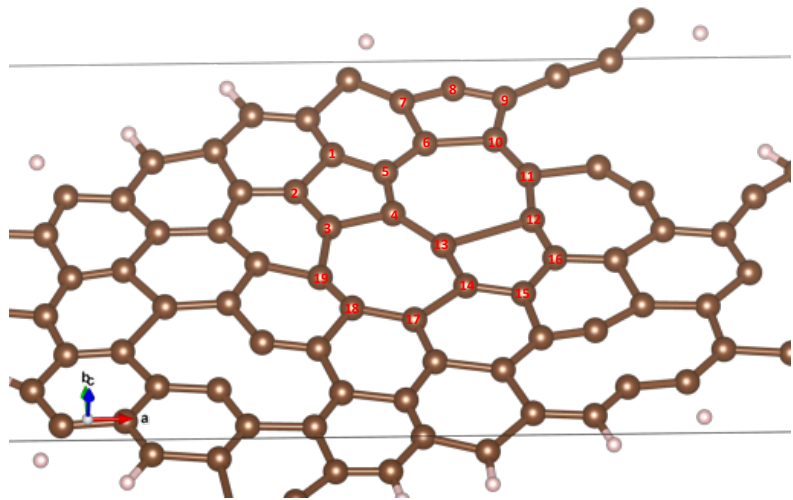 ||||||||
| Temperature (K) | $C_{1-2}$ (Å) | $C_{2-3}$ (Å) | $C_{3-4}$ (Å) | $C_{4-5}$ (Å) | $C_{5-6}$ (Å) | $C_{6-7}$ (Å) | $C_{7-8}$ (Å) |
| 4000 | 1.529 | 1.403 | 1.628 | 1.358 | 1.378 | 1.474 | 1.324 |
| Temperature (K) | $C_{8-9}$ (Å) | $C_{9-10}$ (Å) | $C_{10-11}$ (Å) | $C_{11-12}$ (Å) | $C_{12-13}$ (Å) | $C_{13-14}$ (Å) | $C_{6-10}$ (Å) |
| 4000 | 1.270 | 1.390 | 1.411 | 1.470 | 2.336 | 1.413 | 1.693 |
| Temperature (K) | $C_{14-15}$ (Å) | $C_{15-16}$ (Å) | $C_{14-17}$ (Å) | $C_{17-18}$ (Å) | $C_{18-19}$ (Å) | $C_{19-3}$ (Å) | $C_{13-4}$ (Å) |
| 4000 | 1.412 | 1.423 | 1.698 | 1.544 | 1.256 | 1.674 | 1.596 |
| Temperature (K) | $C_{16-12}$ (Å) | | | | | | |
| 4000 | 1.351 | | | | | | |

**Table 2**: Distances between C atoms in 5-7-7-5 defect regions as formed at 4000 K temperature after 2 picoseconds.

While the displacement of the atoms in x and y directions are the most and increase as the temperature goes higher, the bond angles between C atoms inside the hexagonal, heptagonal, and pentagonal rings along with angles between the shared atoms belonging to each type of these rings vary between 1-5 degrees dependent on the temperature range up to 1200 K. The distances between C atoms in heptagonal and pentagonal rings present at 300 K and 1200 K temperatures are marked and noted in Table 1. The resulting structures however do not show many changes or reconstructions compared to each other in this temperature range. However, once the system is heated up to a 4000 K, the defects propagate and rearrange themselves to form 5-7-7-5 defects. The bond lengths of the newly formed pentagonal and heptagonal rings are listed in Table 2. The average bond lengths in pentagonal, hexagonal, and heptagonal rings at 300 K are 1.713 Å, 1.443

Å and 1.447 Å, respectively. At 2000 K, only hexagonal rings stay with a changed average bond length of 1.422 Å. As the temperature rises to 4000 K, the system fully reconstructs itself to form the 5-7-7-5 defects with average bond length of the 7-members and 5-members rings to be 1.430 Å and 1.606 Å, respectively.

*Temperature-dependent system evolutions with ad-atoms*

The presence of impurities is quite common in materials in general and could be responsible for altering electronic properties as well as introducing point defects in materials making them suitable for technological applications such as in electronic and optoelectronic devices, gas sensors, biosensors, and batteries for energy storage. For graphene, two types of defects[90-92] are common. One is Stone-Wales that is generated by a pure reconstruction of a graphene lattice (switching between pentagons, hexagons, and heptagons). Here no atoms are added or removed. Previous section has explored this scenario. Another is defect reconstruction that can either originate by removing an atom from its lattice position such that the structure relaxes into a lower energy state with a different bonding geometry or by adding a foreign atom to bridge, hollow or top sites potentially leading to different bonding with the graphene atoms.

As example cases, we have studied two different scenarios of putting (a) transitional metal atom[93] such as Cr and (b) CH, $CH_2$, $CH_3$ groups[94] as ad-atoms at different temperatures. While the first choice is mostly driven by already studied effects of metal-carbon binding effects with direct applications into developing battery materials, the latter is propelled by the self-healing mechanism of graphene sheets observed under electron beams. For both studies, the configurational space to be explored is huge in terms of choosing the initial positions of ad-atoms, distances between the atoms and surfaces, or the bonds to be broken to see if graphene sheet can rearrange itself. This also creates a computational challenge for such a workflow to investigate all possible configurations. A sample averaging technique can be employed to perform such studies.[95] For this project, we have limited ourselves to probing a few representative cases. While for Cr ad-atom we have chosen one configuration for each of the bridge, hollow and top sites, for CH groups, three different initial configurations are constructed by breaking bonds between C, H atoms and placing the molecules on the top. The temperature-dependent dynamics is then explored using AIMD simulations at 300 K.

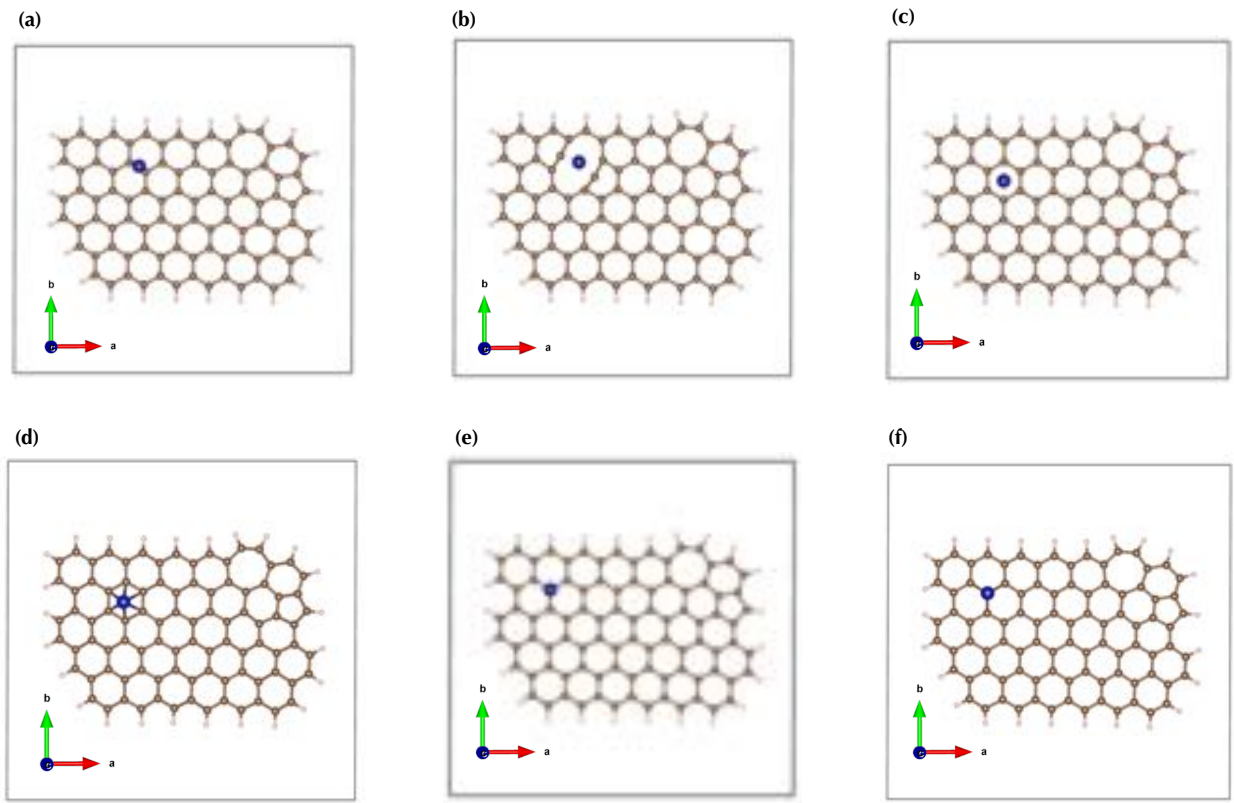

**Figure 5**: **Snapshots from AIMD simulations with Cr ad-atom performed at 300 K**. Initial configurations where Cr atom is added to the bridge (a), hollow (c) and top (e) sites and dynamics is studied at 300 K. The positions after 10 fs (b), 14 fs (d) and 2000 fs (f) are shown in the figure.

Cr atom is added to one of the bridge, hollow and top sites in the graphene cell as demonstrated by Figure 5(a,c,e). Magnetism was not considered for the Cr atom. The ad-atom, initially assigned to the bridge site, likes to diffuse into the system causing bond-breakings between the hexagonal geometries at 300 K after 10 fs as shown in Figure 5(b). For the hollow site, the metal ion forms a stable bond with the nearest C atoms with an average bond length of 2.165 Å and maintains this formation for the entirety of the simulation as depicted in Figure 5 (d). The ad-atom added to the top site (Figure 5(f)) shows the highest displacement in the c direction at 2.237 Å distance from the closest in-plane C atom. The hollow site is the most stable of all if we compare the total energies such as -910.514 eV, -920.305 eV and -923.351 eV for bridge, top and hollow sites, respectively. We do note that spin-polarized computations along with detailed band structures analyses are needed to be performed to further account for the temperature dependence of the adsorption energies which turn out to be above -8 eV in this nonmagnetic picture. The adsorption is the lowest for adsorption at the hollow site which agrees reasonably well with that reported in the literature. The bond distances between C and Cr atoms can also be varied in which case the binding energy tends to increase when the distance is smaller, leading towards chemical adsorption. This relates to the hybridization of $3d$ orbitals of the metal ad-atom (applicable for other transition metal atoms with partially unoccupied *d* states) and *p* states of graphene and the localized behavior of these orbitals.

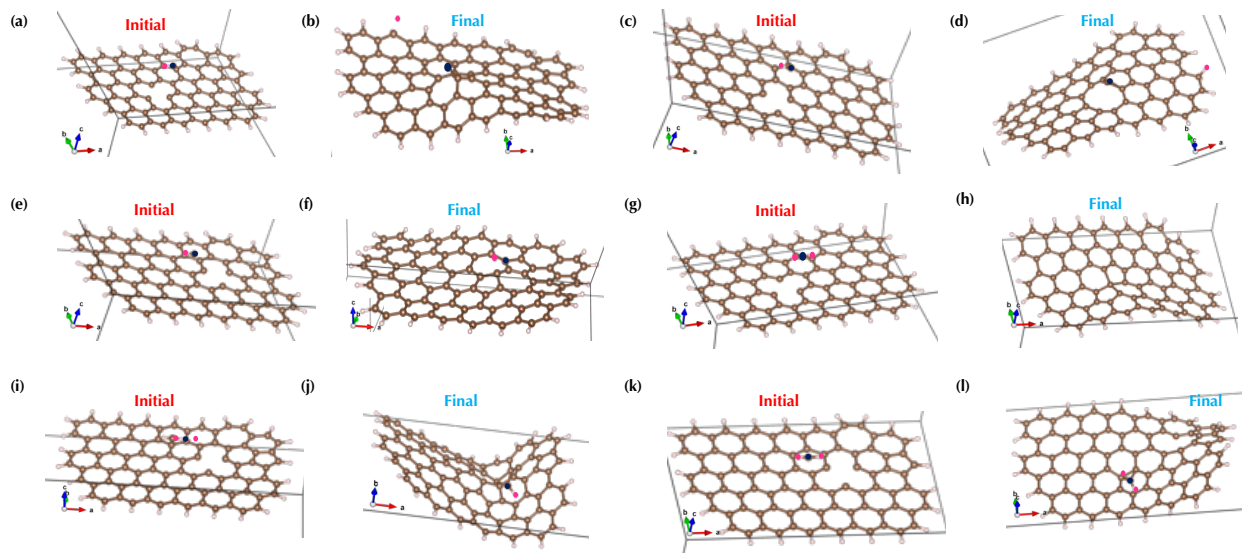

**Figure 6**: **Snapshots from AIMD simulations with CH and CH$_2$ ad-atoms performed at 300 K**. Initial configurations (a,c,e,g,i,k) and final states (b,d,f,h,j,l) where CH and CH$_2$ groups (marked with blue and pink) are added to graphene systems are shown in the figure.

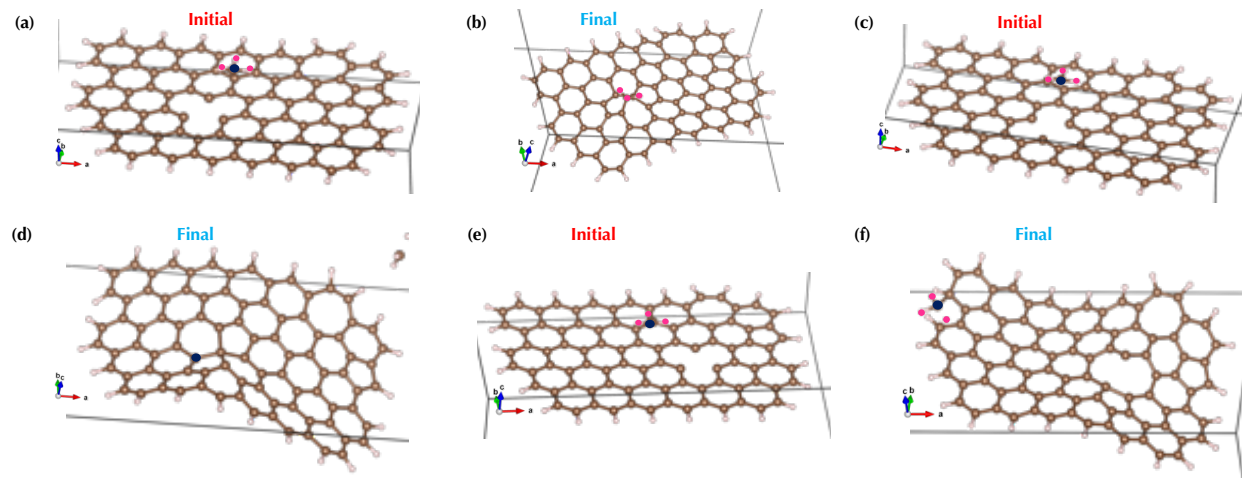

**Figure 7**: **Snapshots from AIMD simulations with CH$_3$ ad-atoms performed at 300 K**. Initial configurations (a,c,e) and final states (b,d,f) where CH$_3$ groups (marked with blue and pink) are added to graphene systems are shown in the figure.

The well-studied healing process of graphene involves C atoms interacting with hole regions, hydrocarbon groups as present as impurities which participate in reconstruction of hexagonal and other rings in the sheet. The results shown by Figures 6 and 7 from AIMD simulations performed at room temperature show the interaction of C atoms with the ad-atoms to reconstruct the geometry. We do see that the holes have healed completely or partially (not all hexagonal rings can be achieved) in all cases depending on the variations in the energy landscapes. Examples of such healings are observed for $CH_3$ molecule in all configurations after 2 ps. The corresponding adsorption energies are -3.709 eV, -7.220 eV and -4.306 eV, respectively for three systems displayed in Figure 6. For CH, $CH_2$ cases, the molecules tend to get attached or bonded to the C atoms at this temperature rather than enough rearrangements required for healing, although interactions of these molecules with the edge atoms can still be observed. Some of the atoms from these hydrocarbons are expected to remain in the environment rather than getting adsorbed to the surface depending on the energy barriers as also seen in the example of $CH_3$ molecule. The adsorption energies for CH and $CH_2$ molecules for these different configurations vary between -7.018 eV and -14.549 eV. From these results, it is also evident that there is a strong dependence of choosing the initial configuration of the bonds and distances between the molecule and the surface along with temperature. This is particularly important and opens avenues to reconnoiter electron beam effects where these mechanisms can be explored under high intensity beam, providing a local heat source to raise the temperature of the graphene surfaces.

### *Regions of interest and end-to-end workflow*

From the above-mentioned studies, it is evident that it is possible to reconstruct geometries utilizing first-principles simulations based on coordinates predicted by a DL framework constructed on microscopic images. This can even be expanded to incorporate edge-computing that involves direct transfer of image-based data from microscopes via light edge devices such as Nvidia Jetson series and then analyze, train DL networks using a GPU-based platform followed by performing simulations with CPU/GPU-based high-performance computing resources and feedback to the human in the loop, altogether on-the-fly, to better guide experiments while learning from theoretical models. Furthermore, the results from the simulations as obtained in this study can be used to choose regions of interest, in terms of studying chemical or physical adsorption, healing mechanisms under different experimental conditions. The fully constructed simulation cell can be sampled through randomly selected patches and can determine the similarity of information carried down to select regions to be investigated in the next runs. It is also possible to improve the predictions with higher uncertainties with such outputs. While DL networks and prediction of coordinates of the system of interest fully use open-source packages, in the simulation stage, it is dependent on the users to choose between the level of theoretical accuracy and corresponding potentials or force-fields to enable the active causal loop. Thus, this framework also adheres to the FAIR[38] principles that is necessary to enable findability, accessibility, interoperability, and reusability of data in scientific domains.

## Conclusions:

In summary, we have established the first step towards direct on-the fly data analytics and experiment augmentation in STEM by DFT and MD models. We accentuate that this vision, while actively discussed in the scientific community is highly non-trivial due to extreme disparity between the time scales experimentally accessible to STEM and amenable to atomistic modeling, as well as fundamentally different latencies of imaging and simulation. Thus, matching the two necessitates both the development of infrastructure linking the microscope data streams to the simulation environment, and solving the coupled challenge of ML selection of regions of interest, simulation-based discovery, ultimately enabling feedback to experiment. Here we have shown how an end-to-end workflow can be constructed to study a few instances of graphene physics starting from a microscopic image to performing simulations with the help of DL. While the simulations performed can be of higher accuracy, these still can show the efficacy of this framework which is dependent on feature predictions using DL from images and using those to provide initial conditions for more theoretical studies.

Overall, this approach allows for a couple of significant advancements in the communities of DL applications to both experiments and simulations. The first is the rapid exploration and analyses of images to extract features with associated uncertainties and a reasonable comparison between these predictions with computational simulations at different length scales. The second is to utilize "the human in the loop" along with the results from observational and synthetic data to improve the DL frameworks for better adaptability, even under different experimental conditions compared to that utilized in training. Finally, we pose that enabling this workflow will allow formulating the specific physical and chemical challenge that will push, but not hopelessly exceed, the regions of experimentally accessible.

**Acknowledgements**:

This effort (machine learning) is based upon work supported by the U.S. Department of Energy (DOE), Office of Science, Office of Basic Energy Sciences Data, Artificial Intelligence Nanoscale Science Research (NSRC AI) Centers program (A.G., BGS, S.V.K.), and was also supported (STEM experiment) by the DOE, Office of Science, Basic Energy Sciences (BES), Materials Sciences and Engineering Division (O.D.), by the DOE, Office of Science, Basic Energy Sciences (BES), and was performed and partially supported (M.Z., B.G.S.) at the Oak Ridge National Laboratory's Center for Nanophase Materials Sciences (CNMS), a DOE Office of Science User Facility.

## Methods:

### Samples preparation and Imaging

Graphene sample used in this work was grown using atmospheric pressure chemical vapor deposition (AP-CVD) and Nion UltraSTEM 200 was used to perform STEM imaging. Associated all other details of the measurements can be found in this reference.[84]

### DFT and AIMD

Details on DFT simulations for the graphene movies: DFT simulations GGA were performed using the projector augmented plane-wave (PAW) method and PAW-PBE potential[96] as implemented in the Vienna ab initio simulation package (VASP).[97-98]

A graphene supercell (patch extracted from the full 2,000 atoms supercell) of 91 C atoms with lattice parameters a = b = c = 25.786 Å, with α = β = γ = 90° was considered as the initial structure for performing the full geometry optimization. All the 2-fold C atoms were terminated with H atoms. The structure optimization was performed by relaxing the atoms steadily toward the equilibrium until the Hellman-Feynman forces are less than $10^{-3}$ eV/Å. All geometry optimization computations were carried out with a 400-eV plane-wave cutoff energy with Monkhorst Pack[99] 2*2*2 k-point meshes. Three different configurations such as (a) fixed perimeter atoms positions, (b) fixed coordinates in a,b or (x and y) directions and (c) fixed all H-atoms, were also considered and subjected to optimization.

Details on AIMD simulations for the graphene movies: Ab-initio quantum-mechanical MD simulations were performed using the projector augmented plane-wave (PAW) method and PAW-PBE potential[96] as implemented in the Vienna ab initio simulation package (VASP).[97-98]

The fully optimized structure (H-terminated) obtained from DFT within GGA was considered as the initial structure for performing all AIMD simulations. All computations were carried out with a 400-eV plane-wave cutoff energy with appropriate Monkhorst Pack[99] k-point meshes, at 300 K, 500 K, 700 K, 900 K and 1200 K temperatures with 2000 timesteps of 1 fs each using a Nose-Hoover thermostat. For exploring the ad-atoms, three different sites such as bridge, hollow and top were considered. Metal ions such as Cr were put into these sites to look at the system evolutions at various temperatures. Another range of AIMD simulations involved adding molecular groups like CH, $CH_2$, $CH_3$ to surfaces to study the temperature-dependent dynamics.

The adsorption energy is calculated using the following equation:

$$E_{adsorbate} = E_{system} + E_{adsorbate} - E_{(system + adsorbate)}$$

### DFTB

Density functional based tight binding (DFTB) and the extended tight binding method, enables simulations of large systems and relatively long timescales at a reasonable accuracy and are considerably faster for typical DFT ab initio[100] We incorporated the DFTB approach (version 21.1) into the developed workflow in a similar fashion as detailed for DFT and AIMD. DFTB allowed directly getting reliable results for the graphene systems (graphene-Si) for the full 2043 atom cell on a timeframe of a few hours. A picture of the optimized structure is given in Figure 1 of Supplementary Material. Likewise, DFTB MD could be performed for many picoseconds on the full system. Thus, we note that this approximate DFT approach can permit study of a more complete material system and to evaluate non-local effects in self-healing, etc.

**Length and timescales for each step of the workflow**

The deep learning models relevant to this work were trained using GPU (Nvidia Tesla K80) as provided by the Google Colab platform. The training time varied depending on the training set size and network architecture. A typical network takes ~0.5 hours to train on ~4,500 (256*256 window sizes) sub-images with a 2-3-3-4-3-3-2 architecture (numbers correspond to convolution layers in each U-Net blocks) of a U-Net model. Generally, we train 20 models in an ensemble. Depending on the availability of GPUs, the model can be trained in a sequential or parallel regime. The feature prediction along with creating a simulation object takes a few seconds. A high GPU RAM is preferred to avoid memory issues during training or performing post-training analyses. For a 91 atoms graphene supercell, a full geometry optimization takes up to 10 CPU hours to converge and a similar timescale is applicable for the temperature dependent AIMD simulations. For geometry optimization of the full 2,043 atoms structure, it takes a few hours to converge. This potentially reduces the overall timeframe for the end-to-end workflow down to something more amendable to feedback during a STEM experiment.

**Data Availability**

The data used for all analysis are available through the Jupyter notebooks located at https://github.com/aghosh92/ELIT .

**Code Availability**

The functions used to simulate structures from DL predictions can be found at https://github.com/pycroscopy/atomai .
The details of training DL networks used in this work are available through the Jupyter notebook located at https://github.com/aghosh92/ELIT .
The python-based implementations to construct simulation objects and perform MD simulations can be found at https://github.com/aghosh92/DCNN_MD .

We typically will initialize multiple independent AtomAI models with different seeds and run them on separate GPUs, combine them into ensemble to get the feature predictions. And import them into CPU environment to perform DFT computations.

**Author Contributions**:

S.V.K., M.Z. and A.G. conceived the project. A.G. has implemented the workflow, performed simulations and all analyses. O.D. obtained STEM data on graphene. M.Z. realized the DL framework via AtomAI and PyTorch libraries. B.G.S. ran MD simulations. A.G., M.Z., and S.V.K., wrote the drafts of the manuscript while B.G.S. and O.D. contributed to the writing.

**Competing Interests**:

The authors declare that there are no competing interests.

# References

1. Pennycook, S. J., The impact of STEM aberration correction on materials science. *Ultramicroscopy* **2017,** *180*, 22-33.
2. Pennycook, S. J., Seeing the atoms more clearly: STEM imaging from the Crewe era to today. *Ultramicroscopy* **2012,** *123*, 28-37.
3. Gerber, C.; Lang, H. P., How the doors to the nanoworld were opened. *Nat. Nanotechnol.* **2006,** *1* (1), 3-5.
4. Barth, C.; Foster, A. S.; Henry, C. R.; Shluger, A. L., Recent Trends in Surface Characterization and Chemistry with High-Resolution Scanning Force Methods. *Adv. Mater.* **2011,** *23* (4), 477-501.
5. Bonnell, D. A.; Garra, J., Scanning probe microscopy of oxide surfaces: atomic structure and properties. *Rep. Prog. Phys.* **2008,** *71* (4), 044501.
6. Varela, M.; Findlay, S. D.; Lupini, A. R.; Christen, H. M.; Borisevich, A. Y.; Dellby, N.; Krivanek, O. L.; Nellist, P. D.; Oxley, M. P.; Allen, L. J.; Pennycook, S. J., Spectroscopic imaging of single atoms within a bulk solid. *Phys. Rev. Lett.* **2004,** *92* (9), 095502.
7. Browning, N. D.; Buban, J. P.; Moltaji, H. O.; Pennycook, S. J.; Duscher, G.; Johnson, K. D.; Rodrigues, R. P.; Dravid, V. P., The influence of atomic structure on the formation of electrical barriers at grain boundaries in SrTiO3. *Appl. Phys. Lett.* **1999,** *74* (18), 2638-2640.
8. Nguyen, D. T.; Findlay, S. D.; Etheridge, J., A menu of electron probes for optimising information from scanning transmission electron microscopy. *Ultramicroscopy* **2018,** *184*, 143-155.
9. Grillo, V.; Karimi, E.; Gazzadi, G. C.; Frabboni, S.; Dennis, M. R.; Boyd, R. W., Generation of Nondiffracting Electron Bessel Beams. *Phys. Rev. X* **2014,** *4* (1), 7.
10. Grillo, V.; Harvey, T. R.; Venturi, F.; Pierce, J. S.; Balboni, R.; Bouchard, F.; Gazzadi, G. C.; Frabboni, S.; Tavabi, A. H.; Li, Z. A.; Dunin-Borkowski, R. E.; Boyd, R. W.; McMorran, B. J.; Karimi, E., Observation of nanoscale magnetic fields using twisted electron beams. *Nat. Commun.* **2017,** *8*, 6.
11. Koh, A. L.; Bao, K.; Khan, I.; Smith, W. E.; Kothleitner, G.; Nordlander, P.; Maier, S. A.; McComb, D. W., Electron Energy-Loss Spectroscopy (EELS) of Surface Plasmons in Single Silver Nanoparticles and Dimers: Influence of Beam Damage and Mapping of Dark Modes. *ACS Nano* **2009,** *3* (10), 3015-3022.
12. Scholl, J. A.; Koh, A. L.; Dionne, J. A., Quantum plasmon resonances of individual metallic nanoparticles. *Nature* **2012,** *483* (7390), 421-U68.
13. Senga, R.; Suenaga, K.; Barone, P.; Morishita, S.; Mauri, F.; Pichler, T., Position and momentum mapping of vibrations in graphene nanostructures. *Nature* **2019,** *573* (7773), 247-+.
14. Idrobo, J. C.; Lupini, A. R.; Feng, T. L.; Unocic, R. R.; Walden, F. S.; Gardiner, D. S.; Lovejoy, T. C.; Dellby, N.; Pantelides, S. T.; Krivanek, O. L., Temperature Measurement by a Nanoscale Electron Probe Using Energy Gain and Loss Spectroscopy. *Phys. Rev. Lett.* **2018,** *120* (9).
15. Roushan, P.; Seo, J.; Parker, C. V.; Hor, Y. S.; Hsieh, D.; Qian, D.; Richardella, A.; Hasan, M. Z.; Cava, R. J.; Yazdani, A., Topological surface states protected from backscattering by chiral spin texture. *Nature* **2009,** *460* (7259), 1106-U64.
16. Heinrich, A. J.; Lutz, C. P.; Gupta, J. A.; Eigler, D. M., Molecule cascades. *Science* **2002,** *298* (5597), 1381-1387.
17. Batzill, M.; Diebold, U., The surface and materials science of tin oxide. *Prog. Surf. Sci.* **2005,** *79* (2-4), 47-154.
18. Acharya, D. P.; Camillone, N.; Sutter, P., CO2 Adsorption, Diffusion, and Electron-Induced Chemistry on Rutile TiO2(110): A Low-Temperature Scanning Tunneling Microscopy Study. *Journal of Physical Chemistry C* **2011,** *115* (24), 12095-12105.
19. Schofield, S. R.; Curson, N. J.; Simmons, M. Y.; Ruess, F. J.; Hallam, T.; Oberbeck, L.; Clark, R. G., Atomically precise placement of single dopants in Si. *Phys. Rev. Lett.* **2003,** *91* (13).

20. Fuechsle, M.; Miwa, J. A.; Mahapatra, S.; Ryu, H.; Lee, S.; Warschkow, O.; Hollenberg, L. C. L.; Klimeck, G.; Simmons, M. Y., A single-atom transistor. *Nat. Nanotechnol.* **2012,** *7* (4), 242-246.
21. Eigler, D. M.; Lutz, C. P.; Rudge, W. E., AN ATOMIC SWITCH REALIZED WITH THE SCANNING TUNNELING MICROSCOPE. *Nature* **1991,** *352* (6336), 600-603.
22. Kalinin, S. V.; Pennycook, S. J., Single-atom fabrication with electron and ion beams: From surfaces and two-dimensional materials toward three-dimensional atom-by-atom assembly. *MRS Bull.* **2017,** *42* (9), 637-643.
23. Markevich, A.; Kurasch, S.; Lehtinen, O.; Reimer, O.; Feng, X. L.; Mullen, K.; Turchanin, A.; Khlobystov, A. N.; Kaiser, U.; Besley, E., Electron beam controlled covalent attachment of small organic molecules to graphene. *Nanoscale* **2016,** *8* (5), 2711-2719.
24. Jiang, N., Electron beam damage in oxides: a review. *Rep. Prog. Phys.* **2016,** *79* (1).
25. Gonzalez-Martinez, I. G.; Bachmatiuk, A.; Bezugly, V.; Kunstmann, J.; Gemming, T.; Liu, Z.; Cuniberti, G.; Rummeli, M. H., Electron-beam induced synthesis of nanostructures: a review. *Nanoscale* **2016,** *8* (22), 11340-11362.
26. Dyck, O.; Kim, S.; Kalinin, S. V.; Jesse, S., Placing single atoms in graphene with a scanning transmission electron microscope. *Appl. Phys. Lett.* **2017,** *111* (11).
27. Dyck, O.; Kim, S.; Jimenez-Izal, E.; Alexandrova, A. N.; Kalinin, S. V.; Jesse, S., Building Structures Atom by Atom via Electron Beam Manipulation. *Small* **2018,** *14* (38).
28. Ziatdinov, M.; Maksov, A.; Kalinin, S. V., Learning surface molecular structures via machine vision. *npj Comput. Mater.* **2017,** *3*.
29. Ziatdinov, M.; Dyck, O.; Jesse, S.; Kalinin, S. V., Atomic Mechanisms for the Si Atom Dynamics in Graphene: Chemical Transformations at the Edge and in the Bulk. *Adv. Funct. Mater.* **2019,** *29* (52), 8.
30. Ziatdinov, M.; Maksov, A.; Kalinin, S. V., Learning surface molecular structures via machine vision. *Npj Computational Materials* **2017,** *3*, 31.
31. Rashidi, M.; Wolkow, R. A., Autonomous Scanning Probe Microscopy in Situ Tip Conditioning through Machine Learning. *Acs Nano* **2018,** *12* (6), 5185-5189.
32. Ziatdinov, M.; Dyck, O.; Maksov, A.; Li, X. F.; San, X. H.; Xiao, K.; Unocic, R. R.; Vasudevan, R.; Jesse, S.; Kalinin, S. V., Deep Learning of Atomically Resolved Scanning Transmission Electron Microscopy Images: Chemical Identification and Tracking Local Transformations. *Acs Nano* **2017,** *11* (12), 12742-12752.
33. Gordon, O. M.; Hodgkinson, J. E. A.; Farley, S. M.; Hunsicker, E. L.; Moriarty, P. J., Automated Searching and Identification of Self-Organized Nanostructures. *Nano Letters* **2020,** *20* (10), 7688-7693.
34. Horwath, J. P.; Zakharov, D. N.; Mégret, R.; Stach, E. A., Understanding important features of deep learning models for segmentation of high-resolution transmission electron microscopy images. *npj Computational Materials* **2020,** *6* (1), 108.
35. Lee, C.-H.; Khan, A.; Luo, D.; Santos, T. P.; Shi, C.; Janicek, B. E.; Kang, S.; Zhu, W.; Sobh, N. A.; Schleife, A.; Clark, B. K.; Huang, P. Y., Deep Learning Enabled Strain Mapping of Single-Atom Defects in Two-Dimensional Transition Metal Dichalcogenides with Sub-Picometer Precision. *Nano Letters* **2020,** *20* (5), 3369-3377.
36. Ghosh, A.; Sumpter, B. G.; Dyck, O.; Kalinin, S. V.; Ziatdinov, M., Ensemble learning and iterative training (ELIT) machine learning: applications towards uncertainty quantification and automated experiment in atom-resolved microscopy. *arXiv preprint arXiv:2101.08449* **2021**.
37. Maxim Ziatdinov, A. G., Tommy Wong, Sergei V Kalinin, AtomAI: A Deep Learning Framework for Analysis of Image and Spectroscopy Data in (Scanning) Transmission Electron Microscopy and Beyond. *arXiv preprint* **2021**.
38. Mons, B., Cameron Neylon, Jan Velterop, Michel Dumontier, Luiz Olavo Bonino da Silva Santos, and Mark D. Wilkinson, Cloudy, increasingly FAIR; revisiting the FAIR Data guiding principles for the European Open Science Cloud. *Information Services & Use* **2017,** *37*, 49-56.

39. Ruoqian Lin, R. Z., Chunyang Wang, Xiao-Qing Yang & Huolin L. Xin, TEMImageNet training library and AtomSegNet deep-learning models for high-precision atom segmentation, localization, denoising, and deblurring of atomic-resolution images. *Scientific Reports* **2021,** *11*.
40. A. Jain, S. P. O., G. Hautier, W. Chen, W.D. Richards, S. Dacek, S. Cholia, D. Gunter, D. Skinner, G. Ceder, K.A. Persson, The Materials Project: A materials genome approach to accelerating materials innovation. *APL Materials* **2013,** *1* (1).
41. D. Hicks, M. J. M., E. Gossett, C. Toher, O. Levy, R. M. Hanson, G. L. W. Hart, and S. Curtarolo, The AFLOW Library of Crystallographic Prototypes: Part 2. *Comp. Mat. Sci.* **2019,** *161*, S1-S1011.
42. J. E. Saal, S. K., M. Aykol, B. Meredig, and C. Wolverton, Materials Design and Discovery with High-Throughput Density Functional Theory: The Open Quantum Materials Database (OQMD). *JOM* **2013,** *65*, 1501-1509.
43. M. J. Mehl, D. H., C. Toher, O. Levy, R. M. Hanson, G. L. W. Hart, and S. Curtarolo, The AFLOW Library of Crystallographic Prototypes: Part 1. *Comp. Mat. Sci.* **2017,** *136*, S1-S828.
44. S. Kirklin, S. J. E. S., B. Meredig, B, A. Thompson, J.W. Doak, M. Aykol, S. Rühl and C. Wolverton. , The Open Quantum Materials Database (OQMD): assessing the accuracy of DFT formation energies. *npj Computational Materials* **2015,** *1*.
45. Scheffler, C. D. a. M., The NOMAD laboratory: from data sharing to artificial intelligence. *J. Phys. Mater* **2019,** *2*, 036001.
46. Anubhav Jain, J. M., Shyam Dwaraknath, Nils ER Zimmermann, John Dagdelen, Matthew Horton, Patrick Huck, Donny Winston, Shreyas Cholia, Shyue Ping Ong, Kristin Persson, *The materials project: Accelerating materials design through theory-driven data and tools*. Springer: 2020; p 1751-1784.
47. Ayana Ghosh, F. R., Serge M Nakhmanson, Jian-Xin Zhu, Machine learning study of magnetism in uranium-based compounds. *Phys. Rev. Mat.* **2020,** *4* (6).
48. Ayana Ghosh, L. L., Kapildev K Arora, Bruno C Hancock, Joseph F Krzyzaniak, Paul Meenan, Serge Nakhmanson, Geoffrey PF Wood, Assessment of machine learning approaches for predicting the crystallization propensity of active pharmaceutical ingredients. *CrystEngComm* **2019,** *21* (8), 1215-1223.
49. Batra, R., "Accurate machine learning in materials science facilitated by using diverse data sources. *Nature* **2021**, 524-525.
50. Botu, V., Rohit Batra, James Chapman, and Rampi Ramprasad, Machine learning force fields: construction, validation, and outlook. *The Journal of Physical Chemistry C* **2017,** *121*, 511-522.
51. Cormac Toher, C. O., David Hicks, Eric Gossett, Frisco Rose, Pinku Nath, Demet Usanmaz, Denise C Ford, Eric Perim, Camilo E Calderon, Jose J Plata, Yoav Lederer, Michal Jahnátek, Wahyu Setyawan, Shidong Wang, Junkai Xue, Kevin Rasch, Roman V Chepulskii, Richard H Taylor, Geena Gomez, Harvey Shi, Andrew R Supka, Rabih Al Rahal Al Orabi, Priya Gopal, Frank T Cerasoli, Laalitha Liyanage, Haihang Wang, Ilaria Siloi, Luis A Agapito, Chandramouli Nyshadham, Gus LW Hart, Jesús Carrete, Fleur Legrain, Natalio Mingo, Eva Zurek, Olexandr Isayev, Alexander Tropsha, Stefano Sanvito, Robert M Hanson, Ichiro Takeuchi, Michael J Mehl, Aleksey N Kolmogorov, Kesong Yang, Pino D'Amico, Arrigo Calzolari, Marcio Costa, Riccardo De Gennaro, Marco Buongiorno Nardelli, Marco Fornari, Ohad Levy, Stefano Curtarolo, *The AFLOW fleet for materials discovery*. Springer International Publishing: 2020.
52. Ghanshyam Pilania, A. G., Steven T Hartman, Rohan Mishra, Christopher R Stanek, Blas P Uberuaga, Anion order in oxysulfide perovskites: origins and implications. *npj Computational Materials* **2020,** *6* (1), 1-11.
53. Kenneth M Merz Jr, R. A., Zoe Cournia, Matthias Rarey, Thereza Soares, Alexander Tropsha, Habibah A Wahab, Renxiao Wang, Method and Data Sharing and Reproducibility of Scientific Results. *Journal of Chemical Information and Modeling* **2020,** *60* (12), 5868-5869.
54. Lihua Chen, G. P., Rohit Batra, Tran Doan Huan, Chiho Kim, Christopher Kuenneth, Rampi Ramprasad, Chen, Lihua, et al. "Polymer informatics: Current status and critical next steps. *Materials Science and Engineering: R: Reports* **2021,** *144*.

55. Meredig, B. e. a., Combinatorial screening for new materials in unconstrained composition space with machine learning. *Phys. Rev. B 89*.
56. Nongnuch Artrith, K. T. B., François-Xavier Coudert, Seungwu Han, Olexandr Isayev, Anubhav Jain, Aron Walsh, Best practices in machine learning for chemistry. *Nature Chemistry* **2021,** *13* (6), 505-508.
57. Olexandr Isayev, A. T., Stefano Curtarolo, *Materials Informatics: Methods, Tools, and Applications*. John Wiley & Sons: 2019.
58. Olexandr Isayev, M. P., Alexander Tropsha Methods, systems and non-transitory computer readable media for automated design of molecules with desired properties using artificial intelligence. 2020.
59. Ramprasad, R., Rohit Batra, Ghanshyam Pilania, Arun Mannodi-Kanakkithodi, and Chiho Kim, Machine learning in materials informatics: recent applications and prospects. *npj Computational Materials* **2017,** *3*, 1-13.
60. Schmidt, J., Marques, M. R., Botti, S. & Marques, M. A, Recent advances and applications of machine learning in solid-state materials science. *npj Computational Materials* **2019,** *5*, 1-36.
61. Snyder, J. C., Rupp, M., Hansen, K., Muller, K.-R. & Burke, K, Finding density functionals with machine learning. *Phys. Rev. Lett.* **2012,** *108*.
62. Stanev, V. e. a., Machine learning modeling of superconducting critical temperature. *npj Computational Materials* **2018,** *4*.
63. Stefano Sanvito, M. Ž., J Nelson, T Archer, C Oses, S Curtarolo, *Machine learning and high-throughput approaches to magnetism*. Springer International Publishing: 2020.
64. Eric Schwenker, V. K., Jinglong Guo, Xiaobing Hu, Qiucheng Li, Mark C Hersam, Vinayak P Dravid, Robert F Klie, Jeffrey R Guest, Maria KY Chan, Ingrained--An automated framework for fusing atomic-scale image simulations into experiments. *arXiv preprint* **2021**.
65. Eric Schwenker, W. J., Trevor Spreadbury, Nicola Ferrier, Oliver Cossairt, Maria KY Chan, EXSCLAIM!--An automated pipeline for the construction of labeled materials imaging datasets from literature. *arXiv preprint* **2021**.
66. Eric J Lingerfelt, A. B., Eirik Endeve, O Ovchinnikov, S Somnath, Jose M Borreguero, N Grodowitz, B Park, RK Archibald, CT Symons, SV Kalinin, OE Bronson Messer, M Shankar, Stephen Jesse, BEAM: A computational workflow system for managing and modeling material characterization data in HPC environments. *Procedia Computer Science* **2016,** *80*, 2276-2280.
67. Jacob Madsen, T. S., The abTEM code: transmission electron microscopy from first principles. *Open Research Europe* **2021,** *1* (24), 24.
68. Ondrej Dyck, K. S., Elisa Jimenez-Izal, Anastassia N. Alexandrova, Sergei V. Kalinin,Stephen Jesse., Building structures atom by atom via electron beam manipulation. *Small* **2018,** *14*.
69. Ondrej Dyck, M. Y., Lizhi Zhang, Andrew R. Lupini, Jacob L. Swett, and Stephen Jesse, Doping of Cr in Graphene Using Electron Beam Manipulation for Functional Defect Engineering. *ACS Applied Nano Materials* **2020,** *3* (11), 10855-10863.
70. Dyck, O., Zhang, L., Yoon, M., Swett, J.L., Hensley, D., Zhang, C., Rack, P.D., Fowlkes, J.D., Lupini, A.R. and Jesse, S., Doping transition-metal atoms in graphene for atomic-scale tailoring of electronic, magnetic, and quantum topological properties. *Carbon* **2021,** *173*, 205-214.
71. Dyck, O., Zhang, C., Rack, P.D., Fowlkes, J.D., Sumpter, B., Lupini, A.R., Kalinin, S.V. and Jesse, S., Electron-beam introduction of heteroatomic Pt–Si structures in graphene. *Carbon* **2020,** *161*, 750-757.
72. Lin, J., Yuyang Zhang, Wu Zhou, and Sokrates T. Pantelides, Structural flexibility and alloying in ultrathin transition-metal chalcogenide nanowires. *ACS nano* **2016,** *10*, 2782-2790.

73. Sang, X., Xufan Li, Wen Zhao, Jichen Dong, Christopher M. Rouleau, David B. Geohegan, Feng Ding, Kai Xiao, and Raymond R. Unocic, In situ edge engineering in two-dimensional transition metal dichalcogenides. *Nature Communications* **2018,** *24*, 1-7.
74. Tibor Lehnert, M. G.-A., Janis Köster, Zhongbo Lee, Arkady V. Krasheninnikov, Ute Kaise, Electron-Beam-Driven Structure Evolution of Single-Layer MoTe2 for Quantum Devices. *ACS Applied Nano Materials* **2019,** *2*, 3262-3270.
75. Zhou, W., Xiaolong Zou, Sina Najmaei, Zheng Liu, Yumeng Shi, Jing Kong, Jun Lou, Pulickel M. Ajayan, Boris I. Yakobson, and Juan-Carlos Idrobo, Intrinsic structural defects in monolayer molybdenum disulfide. *Nano Letters* **2013,** *13* (6), 2615-2622.
76. Hong, J., Hu, Z., Probert, M., Li, K., Lv, D., Yang, X., Gu, L., Mao, N., Feng, Q., Xie, L. and Zhang, J., Exploring atomic defects in molybdenum disulphide monolayers. *Nature Communications* **2015,** *6*, 1-8.
77. Toma Susi, J. K., Demie Kepaptsoglou, Clemens Mangler, Tracy C. Lovejoy, Ondrej L. Krivanek, Recep Zan, Ursel Bangert, Paola Ayala, Jannik C. Meyer, and Quentin Ramasse, Silicon–Carbon Bond Inversions Driven by 60-keV Electrons in Graphene. *Physical Review Letters* **2014,** *113*.
78. Dr. Zhiqing Yang, D. L. Y., Dr. Jaekwang Lee, Dr. Wencai Ren, Prof. Hui-Ming Cheng, Prof. Hengqiang Ye, Prof. Sokrates T. Pantelides, Prof. Stephen J. Pennycook, Dr. Matthew F. Chisholm, Direct Observation of Atomic Dynamics and Silicon Doping at a Topological Defect in Graphene. *Angewandte Chemie* **2014,** *126*, 9054-9058.
79. Jaekwang Lee, W. Z., Stephen J. Pennycook, Juan-Carlos Idrobo & Sokrates T. Pantelides Direct visualization of reversible dynamics in a Si6 cluster embedded in a graphene pore. *Nature Communications* **2013,** *4*.
80. Alex W. Robertson, G.-D. L., Kuang He, Euijoon Yoon, Angus I. Kirkland, and Jamie H. Warner, The Role of the Bridging Atom in Stabilizing Odd Numbered Graphene Vacancies. *Nano Letters* **2014,** *14*, 3972-3980.
81. Alex W. Robertson, G.-D. L., Kuang He, Euijoon Yoon, Angus I. Kirkland, and Jamie H. Warner, Stability and Dynamics of the Tetravacancy in Graphene. *Nano Letters* **2014,** *14*, 1634-1642.
82. Zhengyu He, K. H., Alex W. Robertson, Angus I. Kirkland, Dongwook Kim, Jisoon Ihm, Euijoon Yoon, Gun-Do Lee, and Jamie H. Warner, Atomic Structure and Dynamics of Metal Dopant Pairs in Graphene. *Nano Letters* **2014,** *14*, 3766-3772.
83. Ronneberger, O., Fischer, P. & Brox, T. , U-net: Convolutional networks for biomedical image segmentation. . *Med. Image Comput. Comput. Assist. Interv* **2015,** *9351*, 234–241.
84. Ayana Ghosh, B. G. S., Ondrej Dyck, Sergei V Kalinin, Maxim Ziatdinov, Ensemble learning-iterative training machine learning for uncertainty quantification and automated experiment in atom-resolved microscopy. *npj Computational Materials* **2021,** *7* (1), 1-8.
85. Ishaan Gulrajani, D. L.-P., In Search of Lost Domain Generalization. *arXiv preprint* **2020**.
86. al., A. H. L. e., The Atomic Simulation Environment—A Python library for working with atoms. *Journal of Physics: Condensed Matter* **2017,** *29* (27), 273002.
87. David B. Lingerfelt, T. Y., Anthony Yoshimura, Panchapakesan Ganesh, Jacek Jakowski, Bobby G. Sumpter, Nonadiabatic Effects on Defect Diffusion in Silicon-Doped Nanographenes. *Nano Letters* **2021,** *21*, 236-242.
88. Jamie H. Warner, G.-D. L., Kuang He, Alex. W. Robertson, Euijoon Yoon, and Angus I. Kirkland, Bond Length and Charge Density Variations within Extended Arm Chair Defects in Graphene. *ACS Nano* **2013,** *7*, 9860-9866.
89. Stephen T. Skowron, I. V. L., Andrey M. Popovc and; Bichoutskaia, E., Energetics of atomic scale structure changes

in graphene. *Chem. Soc. Rev.* **2015,** *44*.

90. Aoki, H., and Mildred S. Dresselhaus, *Physics of graphene*. Springer Science & Business Media: 2013.
91. Banhart, F., Jani Kotakoski, and Arkady V. Krasheninnikov., Structural defects in graphene. *ACS nano* **2011,** *5* (1), 26-41.
92. Liu, L., Miaoqing Qing, Yibo Wang, and Shimou Chen, Defects in graphene: generation, healing, and their effects on the properties of graphene: a review. *Journal of Materials Science & Technology* **2015,** *31* (6), 599-606.
93. Dyck, O., Yoon, M., Zhang, L., Lupini, A. R., Swett, J. L., & Jesse, S., Doping of Cr in graphene using electron beam manipulation for functional defect engineering. *ACS Applied Nano Materials* **2020,** *3*, 10855-10863.
94. Nakada, K., and Akira Ishii, DFT calculation for adatom adsorption on graphene. *Graphene Simulation* **2011**, 1-19.
95. Lívia B. Pártay*†, A. P. B., and Gábor Csányi, Efficient sampling of atomic configurational spaces. *J. Phys. Chem. B* **2010,** *114* (32), 10502-10512.
96. G. Kresse and G, D. J., From ultrasoft pseudopotentials to the projector augmented-wave method. *Phys. Rev. B* **1999,** *59* (3).
97. Furthmuller, G. K. a. J., Efficient iterative schemes for ab initio total-energy calculations using a plane-wave basis set. *Phys. Rev. B* **1996,** *54* (16).
98. Hafner, G. K. a. J., Ab initio molecular dynamics for liquid metals. *Phys. Rev. B* **1993,** *47* (1).
99. Pack, J. H. M. a. J. D., Special points for Brillouin-zone integrations. *Phys. Rev. B* **1976,** *13* (12).
100. Hourahine, B., Aradi, B., Blum, V., Bonafé, F., Buccheri, A., Camacho, C., Cevallos, C., Deshaye, M.Y., Dumitrică, T., Dominguez, A. and Ehlert, S., DFTB+, a software package for efficient approximate density functional theory based atomistic simulations. *The Journal of Chemical Physics* **2020,** *152*.